
\input harvmac

 1
 1
 1

\font\tenbf=cmbx10
\font\tenrm=cmr10
\font\tenit=cmti10

\font\eightrm=cmr8
\font\eightit=cmti8

\parindent=1.2pc
\magnification=\magstep1
\hsize=6.0truein
\vsize=8.6truein

\def\np#1#2#3{{\it Nucl. Phys.} {\bf B#1} (#2) #3}
\def\pl#1#2#3{{\it Phys. Lett.} {\bf #1B} (#2) #3}

\def\physrev#1#2#3{{\it Phys. Rev.} {\bf D#1} (#2) #3}

\def\prep#1#2#3{{\it Phys. Rep.} {\bf #1} (#2) #3}

\def\Tr{{\rm Tr}~}
\def\ev#1{\langle#1\rangle}
\font\litfont = cmr6

\def\half{{\litfont {1 \over 2}}}

\rightline{hep-th/9506084, RU-95-40, IASSNS-HEP-95/48 }
\vglue0.3cm

\centerline{\tenbf PHASES OF $N=1$ SUPERSYMMETRIC GAUGE THEORIES}
\baselineskip=18pt
\centerline{\tenbf AND ELECTRIC-MAGNETIC TRIALITY\footnote{*}{To appear
in the Proc. of Strings 95.}}
\baselineskip=18pt
\centerline{\eightrm K. INTRILIGATOR$^1$ and N. SEIBERG$^{1,2}$}
\baselineskip=18pt
\centerline{\eightit $^1$Department of Physics, Rutgers University}
\baselineskip=10pt
\centerline{\eightit Piscataway, NJ 08855-0849, USA}
\baselineskip=12pt
\centerline{\eightit $^2$Institute for Advanced Study}
\baselineskip=10pt
\centerline{\eightit Princeton, NJ 08540, USA}

\vglue0.4cm
\centerline{\eightrm ABSTRACT}
\vglue0.2cm
{\rightskip=3pc
 \leftskip=3pc
 \eightrm\baselineskip=10pt\noindent
We discuss the phases of four dimensional gauge theories and demonstrate
them in solvable examples.  Some of our simple examples exhibit
confinement and oblique confinement.  The theory has dual magnetic and
dual dyonic descriptions in which these phenomena happen at weak
coupling. Combined with the underlying electric theory, which gives a
weak coupling description of the Higgs phase, we have
electric-magnetic-dyonic triality.  In an appendix we clarify some
points regarding the use of 1PI superpotentials in these theories.

\vglue0.6cm}

\tenrm\baselineskip=13pt
\leftline{\tenbf 1. Introduction -
the phases of four dimensional gauge theories}
\vglue0.4cm

Recently, it has become clear that certain aspects of four dimensional
supersymmetric field theories can be analyzed exactly, thus providing a
laboratory for the analysis of the dynamics of gauge theories (for a
recent elementary presentation and a list of references see
\ref\powerd{N. Seiberg, hep-th/9506077,
RU-95-37, IASSNS-HEP-95/46, to appear in the
Proc. of PASCOS 95 and in the Proc. of the Oskar Klein lectures.}).
For example, the phases of gauge theories and the mechanisms for phase
transitions can be explored in this context.  The dynamical mechanisms
explored are standard to gauge theories and thus the insights obtained
are expected to also be applicable for non-supersymmetric theories.
Here we will focus on some of these insights.  We summarize the
ideas of
\nref\swi{N. Seiberg and E. Witten, hep-th/9407087,
\np{426}{1994}{19}.}%
\nref\intse{K. Intriligator and N. Seiberg, hep-th/9408155,
\np{431}{1994}{551}.}%
\nref\sem{N. Seiberg, hep-th/9411149 , \np{435}{1995}{129}.}%
\nref\isson{K. Intriligator and N. Seiberg, hep-th/9503179, to appear in
Nucl. Phys. B.}%
\refs{\swi - \isson} and demonstrate them in simple examples.

A gauge invariant order parameter which characterizes the phases of
gauge theories is the Wilson loop:
\eqn\wl{W_w={\rm Tr}_r~Pe^{i\int A}.}
When the loop is a rectangle of length $T$ and width $R$, it has the
following physical interpretation.  Two electrically charged
sources in the representations $r$ and $\bar r$ of the gauge group are
created a distance $R$ apart.  They then propagate for time $T$ when they
are annihilated.  We can use the expectation value
\eqn\limwl{\lim_{T \rightarrow \infty} \ev{W_w}=e^{-TV(R)},}
to find the potential, $V(R)$, between the sources.  The various phases
are characterized by the large $R$ behavior of the potential which, up
to a non-universal additive constant, is
\eqn\wlp{\eqalign{\hbox{Coulomb}\qquad V(R)& \sim {1\over
R}\cr \hbox{free electric}\qquad V(R)& \sim {1\over
R\log(R\Lambda)}\cr
\hbox{free magnetic}\qquad V(R) & \sim {\log(R\Lambda)\over R}\cr
\hbox{Higgs}\qquad V(R)& \sim constant \cr
\hbox{confining}\qquad V(R)& \sim  \sigma R,\cr}}
where $\sigma $ is the string tension.  The free electric phase happens
when the theory has massless photons and electrons.  Then, the electric
charge is renormalized to zero at long distances and leads to the factor
$\log R\Lambda$ in the potential.  Similar behavior occurs when the long
distance theory is a non-Abelian theory which is not asymptotically
free.  The free magnetic phase occurs when there are massless
magnetic monopoles which renormalize the electric coupling constant to
infinity (the $\log R\Lambda$ in the numerator) at large distance.

For a general loop shape, in the Higgs phase the Wilson loop has
exponential falloff in the the perimeter of the loop and in the
confining phase there is exponential falloff in the area of the loop.

In addition to the familiar Abelian Coulomb phase, there are theories
which have a non-Abelian Coulomb phase with massless interacting quarks
and gluons exhibiting the above Coulomb potential.  This phase occurs
when there is a non-trivial, infrared fixed point of the renormalization
group.  These are thus non-trivial, interacting four dimensional
conformal field theories.

Another order parameter is the 'tHooft loop $W_t$ constructed by cutting
a loop out of the space and considering non-trivial (twisted) boundary
conditions around it.  In a fashion similar to the Wilson loop, it can
be interpreted as creating and annihilating a monopole anti-monopole
pair.  The potential between the monopoles, obtained from the 'tHooft
loop via $\lim_{T \rightarrow \infty} \ev{W_t}=e^{-TV(R)}$, satisfies for
large $R$
\eqn\tlp{\eqalign{\hbox{Coulomb}\qquad V(R)& \sim{1\over R}\cr
\hbox{free electric}\qquad V(R)&\sim {\log(R\Lambda)\over R}\cr
\hbox{free magnetic}\qquad V(R)& \sim{1\over R\log(R\Lambda)}\cr
\hbox{Higgs}\qquad V(R)& \sim \rho R \cr
\hbox{confining}\qquad V(R)& \sim constant \cr}}
up to an additive non-universal constant.  The linear potential in the
Higgs phase reflects the string tension in the Meissner effect.

Note that in going from the Wilson loop to the 'tHooft loop the
behavior in the free electric and the free magnetic phases are
exchanged.  This is a reflection of the fact that under
electric-magnetic duality, which exchanges electrically charged fields
with magnetically charged fields, the Wilson loop and the 'tHooft loop
are exchanged.  Mandelstam and 'tHooft suggested that, similarly,
the Higgs and confining phases are exchanged by duality.  Confinement
can thus be understood as the dual Meissner effect associated with a
condensate of monopoles.

Dualizing a theory in the Coulomb phase, we remain in the same phase (the
behavior of the potential is unchanged).  For an Abelian Coulomb phase
with free massless photons this follows from a standard duality
transformation.  What is not obvious is that this is also the case in
a non-Abelian Coulomb phase.  This was first suggested by
Montonen and Olive
\ref\mo{C. Montonen and D. Olive, \pl {72}{1977}{117}; P. Goddard,
J. Nuyts and D. Olive, \np{125}{1977}{1}.}.
The simplest version of their proposal is
true only in $N=4$ supersymmetric field theories
\ref\dualnf{H. Osborn, \pl{83}{1979}{321}; A. Sen, hep-th/9402032,
\pl{329}{1994}{217}; C. Vafa and E. Witten, hep-th/9408074,
\np{432}{1994}{3}.}
and in finite $N=2$ supersymmetric theories
\ref\swii{N. Seiberg and E. Witten, hep-th/9408099,
\np{431}{1994}{484}.}.  The extension of these ideas to
asymptotically free $N=1$ theories appeared in \sem.

Another order parameter is the product $W_d=W_wW_t$, which corresponds to a
dyon loop.  Making a table with the dependence of the order parameters
in the phases suggests an ``oblique'' confinement phase
\nref\thooft{G. 'tHooft, \np{190}{1981}{455}.}%
\nref\cardyrabin{J. Cardy and E. Rabinovici, \np{205}{1982}{1};
J. Cardy, \np{205}{1982}{17}.}%
\refs{\thooft,\cardyrabin}
$$\vbox{\rm \settabs 4 \columns
\+ & $W_w$ & $ W_t$ & $W_d$ \cr
\+ Higgs & perimeter & area & area \cr
\+ Confinement& area& perimeter & area\cr
\+ Oblique Conf.& area & area & perimeter \cr}$$
Whereas the Higgs phase is associated with an electrically charged
condensate, the confining phase can be associated with a condensate of
monopoles and the oblique confinement phase can be associated with a
condensate of dyons.

In $SU(N_c)$ theories with matter in the fundamental representation, the
elementary quarks can screen the charges involved in the above loops and
thus all loops have perimeter behavior.  Indeed, there is no distinction
between Higgs and confinement in these theories
\ref\fund{T. Banks and E. Rabinovici, \np{160}{1979}{349}; E. Fradkin
and S. Shenker, \physrev{19}{1979}{3682}.}.
This suggests consideration of gauge theories with matter not in the
fundamental representation of the gauge group.  More precisely, we need
matter fields in a non-faithful representation of the center of the
gauge group.

In the next two sections we will discuss an $SU(2)$ gauge theories with
matter $Q$ in the adjoint representation.  It is then possible to study
confinement by considering Wilson loops in the fundamental
representation of $SU(2)$.  The quarks in the adjoint representation are
unable to screen the $Z_2$ center of the gauge group.

In the confining phase there is often a mass gap with no massless
particles (or the massless particles are free).  In that case the 1PI
effective Lagrangian for operators does not suffer from infrared
divergences.  The superpotential of this effective action can be
obtained following techniques discussed in
\nref\vy{G. Veneziano and S. Yankielowicz \pl{113}{1982}{231};
T. Taylor, G. Veneziano and Yankielowicz, \np{218}{1983}{439}.}%
\nref\cern{D. Amati, K. Konishi, Y. Meurice, G.C. Rossi and G.
Veneziano, \prep{162}{1988}{169} and references therein.}%
\nref\kaplou{V. Kaplunovsky and J. Louis, \np{422}{1994}{57}.}%
\nref\ils{K. Intriligator, R.G. Leigh and N. Seiberg, hep-th/9403198,
\physrev{50}{1994}{1092}; K. Intriligator, hep-th/9407106,
\pl{336}{1994}{409}.}%
\nref\iipi{C.P. Burgess, J.-P. Derendinger, F. Quevedo, and M. Quiros,
hep-th/9595171, CERN-Th/95-111.}%
\refs{\vy - \iipi}.  In an appendix we will present our understanding of
this effective action and its proper use.  In particular, we will show
when it leads to incorrect conclusions when interpreted as a Wilsonian
effective action.

\vglue0.6cm
\leftline{\tenbf 2. $SU(2)$ with one adjoint, $Q$; an Abelian Coulomb phase}
\vglue0.4cm

This is the $N=2$ theory discussed in \swi.  The theory has a quantum
moduli space of vacua labeled by the expectation value of the massless
meson field $M= Q^2$.  The $SU(2)$ gauge symmetry is broken to $U(1)$ on
this moduli space, so the theory has a Coulomb phase with a massless
photon.

As discussed in \swi, there is a massless magnetic monopole field
$q_{(+)}$, at $M=4\Lambda ^2$ and a massless dyon $q_{(-)}$ at\foot{We
use the conventions of
\nref\finnpou{D. Finnell and P. Pouliot, RU-95-14, SLAC-PUB-95-6768,
hep-th/9503115.}%
\refs{\ils, \finnpou} where the normalization of $\Lambda^2$ (in the
$\overline{DR}$ scheme) differs by a factor of 2 from that of \swi; our
order parameter $M$ is related to $u$ of \swi\ as $u=\half M$.}
$M=-4\Lambda ^2$.  Therefore, these two points are in a free magnetic
and a free dyonic phase, respectively.  Here $q_{(+)}$ is a doublet
charged under the magnetic $U(1)_M$, which is related to the electric
$U(1)_E$ by the electric-magnetic transformation $S$: $F\rightarrow
\tilde F$ (modulo $\Gamma(2) \subset SL(2,Z)$).  Similarly, $q_{(-)}$ is
a doublet charged under a dyonic $U(1)_D$, related to $U(1)_E$ by the
$SL(2,Z)$ transformation $ST$ (again, modulo $\Gamma(2) \subset
SL(2,Z)$), where $T$ is a rotation of the theta angle by $2\pi$.  Near
where these fields are massless, they couple through the effective
superpotentials
\eqn\mdwii{W_\pm \sim  \left(M  \mp 4\Lambda ^2\right)
q_{(\pm )}\cdot q_{(\pm )}.}

Referring to the underlying $SU(2)$ theory as ``electric,'' we can say
that it has two dual theories.  One of them, which we can refer to as
the ``magnetic dual,'' describes the physics around $M=4\Lambda^2$ with
the superpotential $W_+$.  The other dual, which can be called the
``dyonic dual,'' is valid around $M=-4\Lambda^2$ and is described by
$W_-$.  Consider
giving $Q$ a mass by adding a term $W_{tree}= \half mM$ in
the electric theory.  Adding $W_{tree}$ to \mdwii, the equations of
motion give $\ev{q_{(\pm)}\cdot q_{(\pm)}}\sim m$ and lock $\ev{M}=\pm
4\Lambda ^2$.  The condensate of monopoles/dyons Higgses the dual theory
and thus gives confinement/oblique confinement of the electric theory
by the dual Meissner effect \swi.
\eject
\leftline{\tenbf 3.
$SU(2)$ with two adjoints; A non-Abelian Coulomb phase}
\vglue0.4cm
\leftline{\tenit 3.1 The ``electric'' theory}
\vglue0.4cm

This theory has $N=1$ (not $N=2$) supersymmetry.  Writing the matter
fields as $Q^i$ with $i=1,2$ a flavor index, there is a 3 complex
dimensional moduli space of classical vacua parametrized by the
expectation values of the gauge singlet fields $M^{ij}=Q^i\cdot Q^i$.
In the generic vacuum $\ev{Q^1}$ breaks $SU(2)$ to a $U(1)$ which is
then broken by $\ev{Q^2}$.  For $\det \ev{M^{ij}}\neq 0$, the gauge
group is completely broken and the theory is in the Higgs phase.  On the
non-compact two complex dimensional subspace of vacua with $\det M=0$,
there is an unbroken $U(1)$ gauge symmetry and thus a light photon along
with a pair of massless electrically charged fields. At the point
$\ev{M}=0$ the $SU(2)$ gauge group is unbroken.

We now turn to the quantum theory.  The theory has the global symmetry
group $SU(2)\times U(1)_R$, with $Q$ transforming as ${\bf 2}_{\half}$,
which determines that any dynamically generated superpotential must be
of the form
\eqn\wnfiiwsym{W={c\over \Lambda}\det M,}
with $c$ a dimensionless constant.  Its behavior at $M\rightarrow
\infty$ is incompatible with asymptotic freedom, as signaled by the
presence of the scale $\Lambda$ in the denominator.  Therefore, no
superpotential can be generated and the classical vacuum degeneracy
outlined above is not lifted quantum mechanically.

The generic ground state with generic $M$ is in the Higgs phase.
Consider now the subspace of the moduli space with $\det M=0$.  The low
energy degrees of freedom there are a single photon, a pair of massless
electrically charged fields and some neutral fields.  This theory cannot
become strong in the infrared.  In fact, the loops of the massless
charged fields renormalize the electric charge to zero.  Therefore, this
subspace of the moduli space is in a free electric phase.

Now consider adding a tree level superpotential $W_{tree}=\half \Tr mM$.
Taking $m=\pmatrix{0&0\cr 0&m_{2}}$, $Q^2$ gets a mass and can be
integrated out.  The low energy theory is $SU(2)$ with a single massless
adjoint matter field, which is the example of the previous section.  Its
scale $\Lambda_L$ can be expressed in terms of the scale $\Lambda$ of
the high energy theory and the mass as $\Lambda_L^4 = m_2^2 \Lambda^2$.
Therefore, the massless monopole and dyon are at $\ev{M^{11}}=\pm
4m_{2}\Lambda$.  Note that as $m_2\rightarrow 0$ the point $\ev{M}=0$
has both massless monopoles and dyons.  These are mutually
non-local\foot{ A similar situation was found in $N=2$ $SU(3)$ Yang
Mills theory
\ref\AD{P. Argyres and M. Douglas, hep-th/9505062, IASSNS-HEP-95-31.}.}
and signal another phase at this point in the theory with $m_2=0$.  We
interpret this as a non-Abelian Coulomb phase \intse.

Starting from the theory with $m_2\neq 0$, turning on $m_1\neq 0$ drives
the monopole or dyon to condense and the vacuum is locked at
$\ev{M^{11}}=\pm 4m_2\Lambda$.  The $+$ sign is a vacuum with monopole
condensation and thus confinement.  The $-$ sign is a vacuum with dyon
condensation and thus oblique confinement.  More generally, these vacua
are at $\ev{M^{ij}}=\pm 4\Lambda \det m(m^{-1})^{ij}$.  These
expectation values can be obtained from
\eqn\we{W_e={e\over 8\Lambda}\det M+\half \Tr mM,}
with $e=\mp 1$ for confinement and oblique confinement, respectively.

The theory has various phase branches.  For mass $m=0$ there is a Higgs
phase which, in terms of $W_e$, corresponds to $e=0$.  There is a
subspace $\det M=0$ in the free electric phase and the point $M=0$ in
a non-Abelian Coulomb phase.  For $m\neq 0$ but with $\det m=0$ the
theory is in the Coulomb phase with a free magnetic point and a free
dyonic point.  For $\det m\neq 0$ the theory is either confining and
described by the superpotential with $e=-1$ or it is oblique confining and
described by the superpotential with $e=1$.

Note that the theory has three branches.  Every one of them has its own
superpotential ($e=0,1,-1$ in \we).  We will return to the meaning of
this superpotential and how it could be different in the various phases
in the appendix.

\vglue0.6cm
\leftline{\tenit 3.2 Dual non-Abelian theories}
\vglue0.4cm

The analysis of \isson\ reveals that this theory has two dual
theories, labeled by $\epsilon =\pm 1$.  The two theories are based on
an $SU(2)$ gauge group with two fields $q_i$ in its adjoint representation
and three gauge singlet fields $M^{ij}$.  The difference
between the two theories is in the superpotential
\eqn\wepe{W_{\epsilon} ={1\over 12\sqrt{\Lambda \tilde \Lambda}}
M^{ij}q_i\cdot q_j+\epsilon \left({1\over 24\Lambda}\det M+{1\over 24\tilde
\Lambda}\det q_i\cdot q_j\right),}
where $\tilde \Lambda$ is the scale of the dual $SU(2)$.  Here the
elementary field $M^{ij} $ was rescaled to have dimension 2 just as its
counterpart $M^{ij}= Q^i\cdot Q^j$ in the electric theory.  The theory
with $\epsilon =1$ is a ``magnetic'' dual and that with $\epsilon =-1$ a
``dyonic'' dual.

We now analyze the dynamics of these dual theories.  Since they are
similar to the theory studied in the previous subsection, we proceed as
we did there. These theories have three phases: Higgs, confining and
oblique confining.  We study them using the gauge invariant order
parameters $N_{ij}\equiv q_i\cdot q_j$.  Its effective superpotential is
obtained by writing the tree level superpotential \wepe\ in terms of $N$
and adding to it ${\tilde e \over 8 \tilde \Lambda} \det N$ where, in
the Higgs, confining and oblique confinement branches, $\tilde e= 0, -1,
1$, respectively
\eqn\wepen{W_{\epsilon, \tilde e} ={1\over 12\sqrt{\Lambda \tilde
\Lambda}}\Tr MN+\epsilon \left({1\over 24\Lambda}\det M+{1\over 24\tilde
\Lambda}\det N\right)+{\tilde e\over 8\tilde \Lambda}\det N.}
Now we can integrate out the massive field $N$ to find
\eqn\wepenon{W_{\rm eff} ={1\over 8\Lambda}{\tilde e-\epsilon\over
1+3\tilde e\epsilon}\det M.}
This is the same as the effective superpotential \we\ of the electric
theory with
\eqn\etran{e={\tilde e-\epsilon\over 1+3\tilde e\epsilon}.}
We see that the various phases are permuted in the different
descriptions as:
\smallskip
$$\vbox{\rm \settabs 4 \columns
\+ {\bf Theory}\qquad\qquad\qquad &\ &{\bf Phases}&\ \cr
\+ &\ &\ &\ \cr
\+ electric & Higgs ($e=0$) & conf. ($e=-1$) & obl. conf. ($e=1$) \cr
\+ magnetic ($\epsilon=1$) & obl. conf. ($\tilde e=1$) &  Higgs ($\tilde
e=0$ ) & conf. ($\tilde e=-1$) \cr
\+ dyonic ($\epsilon =-1$) & conf. ($\tilde e=-1$) & obl. conf. ($\tilde
e=1$) & Higgs ($\tilde e=0$) \cr} $$

\smallskip

It is a simple exercise to check that by dualizing the magnetic and
dyonic theories as we above dualized the electric theory (two duals of
each), we find permutations of the same three theories. The $S_3$
triality permuting the phases and branches is associated with a quotient
of the $SL(2,Z)$ electric-magnetic duality symmetry group: the theories
are preserved under $\Gamma (2)\subset SL(2,Z)$, leaving the quotient
$S_3=SL(2,Z)/\Gamma (2)$ with a non-trivial action.

This discussion leads to a new interpretation of the first term in \we .
In the electric theory this term appears as a consequence of complicated
strong coupling dynamics in the confining and the oblique confinement
branches of the theory.  In the dual descriptions it is already present
at tree level.

Consider the theory with a mass $m_2$ for $Q^2$. As discussed above, the
low energy electric theory has a Coulomb phase with massless monopoles
or dyons at the strong coupling singularities $\ev{M^{11}}=\pm
4m_2\Lambda$.  We now derive this result in the dual theories.  Adding
$W_{tree}=\half m_2M^{22}$ to the superpotential \wepe\ of the dual
theory, the equations of motion give
\eqn\dceom{\eqalign{{1 \over 12\sqrt{\Lambda \tilde \Lambda}
} q_2 \cdot q_2 +{8\epsilon \over 24\Lambda} M^{11}+{1\over 2}m_2&=0
\cr q_1 \cdot q_2&=0}\qquad
\eqalign{M^{22}&=-\half \epsilon\sqrt{\Lambda \over \tilde \Lambda}
q_1 \cdot q_1 \cr M^{12}&=0.}}
For $q_2^2\neq 0$, $\ev{q_2}$ breaks the gauge group to $U(1)$ and the
remaining charged fields $q_1^{\pm}$ couple through the low
energy superpotential
\eqn\dcloww{{1\over 16\sqrt{\Lambda \tilde \Lambda}}
(M^{11}-4\epsilon m_2 \Lambda)q_1^+q_1^-.}
(This superpotential is corrected by contributions from instantons in
the broken magnetic $SU(2)$ theory.  However, these are negligible near
$M^{11}=4\epsilon m_2 \Lambda$.)  We see that the theory has a charged
doublet of massless fields $q_1^\pm$ at $M^{11}=4\epsilon m_2\Lambda$,
exactly as expected from the analysis of the electric theory.  There
these states appeared as a result of strong coupling effects.  Here we
see them as weakly coupled states in the
dual theories.  This is in accord with the interpretation of the
$\epsilon=1$ ($\epsilon=-1$) theory as magnetic (dyonic).

The other monopole point on the moduli space of the theory with $m_1=0$
but $m_2\not=0$ is at $M^{11}=-4\epsilon m_2\Lambda$.  It
arises from strong coupling dynamics in the dual
theories.  To see that, note that the above analysis is not valid when
the expectation value of $q_2$ is on the order of or smaller than the
mass of $q_1$.  In that case, $q_1$ should be integrated out first.  The
equations of motion in the low energy theory yield a single massless
monopole point at $M^{11}=-4\epsilon m_2\Lambda $ \isson.

An analysis similar to the one above leads to a strongly coupled
state in the dual theories along the flat directions with $\det M=0$
in the $m=0$ case.  This state can be interpreted as the massless
quark of the electric theory in that free electric phase.

To conclude, this theory has three branches which are in three different
phases: Higgs, confining and oblique confinement (various submanifolds of
these branches are in Coulomb, free electric, free magnetic and free
dyonic phases).  They touch each other at a point in a non-Abelian
Coulomb phase.  Corresponding to the three branches there are three
different Lagrangian descriptions of the theory: electric, magnetic and
dyonic.  Each of them describes the physics of one of the branches, where
it is Higgsed, in weak coupling and the other two in strong coupling.

In both the example of the previous section and this one, the theory has a
discrete symmetry which relates the confining and the oblique
confinement phases\foot{This symmetry is manifest only in the electric
description.  In the dual descriptions it is realized as a quantum
symmetry \isson.}.  Therefore, in these cases the effects of confinement
are indistinguishable from the effects of oblique confinement.
Correspondingly, the magnetic and the dyonic descriptions are similar --
they differ only in the sign of $\epsilon$.  In other examples,
discussed in \isson, these two phases are not related by a symmetry and
the two dual descriptions look totally different.

\vglue0.6cm
\vbox{\leftline{\tenbf Appendix: The superpotential in the confining phase,
1PI effective}
\leftline{\tenbf action, Legendre transform and ``integrating in''}}
\vglue0.4cm

There are two different objects which are usually called ``the effective
action:'' the 1PI effective action and the Wilsonian one.  When there
are no interacting massless particles, these two effective actions are
identical.  This is often the case in the confining phase.  However,
when interacting massless particles are present, the 1PI effective
action suffers from IR ambiguities and might suffer from holomorphic
anomalies
\ref\sv{M.A. Shifman and A. I. Vainshtein, \np{277}{1986}{456};
\np{359}{1991}{571}.}.
These are absent in the Wilsonian effective action.

Consider the theory with a tree level superpotential with sources for
the gauge invariant polynomials $X^r$ in the matter fields,
$W_{tree}=\sum _r g_rX^r$, with the $g_r$ regarded as background chiral
superfield sources
\ref\nonren{N. Seiberg, hep-ph/9309335, \pl{318}{1993}{469}.}.
The functional integral with the added source terms gives the
standard generating function for the correlation functions, $\Gamma(g)$.
If supersymmetry is not broken, $\Gamma(g)$ is supersymmetric (otherwise
we should include the Goldstino field and supersymmetry will be realized
non-linearly) and\foot{In writing this expression we should think of the
coupling constants $g_r$ as background superfields.  Otherwise, $W_L(g)$
is a constant superpotential, which has no effect in global
supersymmetry.  Indeed, the following equation can be interpreted as
differentiating the action with respect to the $F$ component of $g_r$.}
$\Gamma (g)=\dots +\int d^2\theta W_L(g)$.  Using $W_L(g)$ we can
compute the expectation values
\eqn\corr{{\partial W_L(g)\over \partial g_r}=\ev{X^r}.}
It is standard to perform a Legendre transform to find the 1PI effective
action for the operators $X_r$:
\eqn\invlt{W_{dyn}(X^r)=\left(W_L(g_r)-\sum _rg_rX^r\right)_{\ev
{g_r}},}
where the $\ev{g_r}$ are the solutions of \corr.  The transformation
{}from $W_L(g_r)$ to $W_{dyn}(X_r)$ can be inverted by the inverse
Legendre transform as
\eqn\wl{W_L(g)=\left(W_{dyn}(X^r)+\sum _rg_rX^r\right) _{\ev{X^r}},}
where the $X^r$ are evaluated at their expectation values $\ev{X^r}$,
which solve
\eqn\eomg{{\partial W_{dyn}\over \partial
X_r}+g_r=0.}

The 1PI effective superpotential
\eqn\weffopi{W_{eff}(X,g)=W_{dyn}(X^r)+\sum _rg_rX^r }
has the property that the equations of motion for the fields $X^r$
derived from it \eomg\ determine their expectation values.  In some
cases the superpotential $W_{eff}$ obtained by the above Legendre
transform is the same as the Wilsonian superpotential for the light
fields.  In applying this procedure we should be careful of the
following pitfalls:

\item{1.} The theory with the sources should have a gap.  Otherwise, the
1PI action is ill defined.

\item{2.}  The theory with the sources might break supersymmetry.
In that case $W_L$ is ill defined.

\item{3.} As the sources are turned off, some particles become massless.
Their interpolating fields should be among the composite fields $X^r$.
If some massless particles cannot be represented by a gauge invariant
operator $X^r$, the effective superpotential derived this way will not
include them.  This often leads to singularities.

\item{4.} The theory might also have other branches (as in the examples
above) which are present only when some sources vanish.  In this case
there are new massless particles at that point and this $W_{eff}$ might
miss some of the branches.  In other words, then the Legendre transform
does not exist.

\item{5.} If some composites do not represent massless particles, they
should be integrated out.  Although we can use the effective
superpotential to find their expectation values, we cannot think of them
as fields corresponding to massive particles except near a point where
they become massless.

When we can use this procedure to find the Wilsonian action, the
linearity of $W_{eff}$ \weffopi\ in the sources provides a derivation of
the linearity of the Wilsonian effective action in the sources.  See
\kaplou\ for a related discussion.

This approach is particularly useful when we know how to compute
$W_L(g_r)$ exactly.  Then, $W_{dyn}$ and $W_{eff}$ follow simply from
the Legendre transform \invlt; this is the ``integration in'' discussed
in \ils .  One situation where $W_L(g_r)$ can be determined is when the
$X^r$ are all quadratic in the elementary fields.  In that case, the
sources $g_r$ are simply mass terms for the matter fields and $W_L(g)$
is the superpotential for the low energy gauge theory with the massive
matter integrated out, expressed in terms of the quantities in the
high-energy theory.

As an example, consider supersymmetric $SU(N_c)$ QCD with $N_f$ flavors.
For $N_f<N_c$ the gauge invariant operators are $M_{i\tilde j}=Q_i\tilde
Q_{\tilde j}$ and their sources are mass terms, $W_{tree}=\Tr m M$.
When the masses are large the quarks can be integrated out.  The low
energy $SU(N_c)$ theory then has a scale $\Lambda _L^{3N_c}=\Lambda
^{3N_c-N_f}\det m$ (again, we use the conventions of \refs{\ils,
\finnpou}).  Gluino condensation in this theory leads to the effective
superpotential for the sources
\eqn\ymw{W_L(m)=N_c(\Lambda ^{3N_c-N_f}\det m)^{1/N_c}.}
Using \invlt, the effective superpotential for the operators $M$ is
\eqn\sqcd{W=(N_c-N_f)\left({\Lambda ^{3N_c-N_f}\over \det
M}\right)^{1/(N_c-N_f)}.}
In this case $W_{dyn}$ agrees with the Wilsonian effective
superpotential
\ref\ads{I. Affleck, M. Dine and N. Seiberg, \np{241}{1984}{493};
\np{256}{1985}{557}.}.

It is also possible to ``integrate in'' more operators which do not
correspond to massless particles.  Then, the effective action can be
used only to compute their expectation values, rather than for studying
them as massive particles.  An example is the ``glueball'' field $S\sim
W_\alpha ^2$, whose source is $\sim \log \Lambda $.  Integrating in $S$
by the Legendre transform of \sqcd\ with the source $\log \Lambda
^{3N_c-N_f}$ yields
\eqn\sqcds{W(S,M)=S\left[\log\left({\Lambda ^{3N_c-N_f}\over
S^{N_c-N_f}\det M}\right)+(N_c-N_f)\right],}
the superpotential obtained in \vy.  Working with such an effective
potential including massive fields can be convenient when interesting
but complicated dynamics is encoded in the integrating out of these
massive fields.  However, as stressed above, we should not think of $S$
as a field describing a massive particle.

As we said above, the analysis of the effective action with sources
might fail to
reveal some of the physics.  For example, for $SU(2)$ with
one adjoint field, discussed in section 2, we can start from the analog of
\ymw\ for this theory,
$W_L(m)=\pm 2(\Lambda ^4m^2)^{1/2}$.  Then equations
\corr\ and \invlt\ give $W=0$ with
the constraint $\ev{M}=\pm 4\Lambda ^2$.  Indeed, adding the source for
$M$ drives the theory to the confining or oblique confining phase with
$\ev{M}=\pm 4\Lambda ^2$. The Coulomb phase cannot be explored in the
theory with a mass term for $Q$.  Similarly, for $SU(2)$ with two
adjoint fields, discussed in section 3, the analog of \ymw\ is
$W_L(m)=\pm 2(\Lambda ^2\det m)^{1/2}$.  Integrating in gives the
confining or oblique confining
phase superpotential \we\ with $e=\pm 1$, missing the $e=0$
phase.  In both of these situations the theory without the sources has
massless particles (the photon and the monopoles or dyons in the theory
of section 2 and the quarks and the gluons in the theory of section 3)
which cannot be represented by the gauge invariant observables.  As we
said above, in a situation like that this method must fail to capture
some of the physics.

Another situation in which the Legendre transform analysis is incomplete
is when supersymmetry is dynamically broken by the added source terms.
A simple example of this is supersymmetric $SU(2)$ with a single field
$Q$ in the $\bf{4}$ of $SU(2)$
\ref\iss{K. Intriligator, N. Seiberg, and S. Shenker,
\pl{342}{1995}{152}.}.
The theory without added source terms has a one complex dimensional
smooth moduli space of vacua labeled by $\ev{X}$, where $X=Q^4$ is the
basic gauge invariant, with a superpotential $W(X)=0$.  Adding a source
$W=gX$ does not lead to a supersymmetric effective superpotential $W(g)$
-- rather, it breaks supersymmetry \iss.  (As discussed in \iss, it is
also possible that there is a non-Abelian Coulomb phase at the origin of
the moduli space and that supersymmetry is unbroken with the added
source term.  In that case the 1PI analysis again fails to capture the
physics.)

\vglue0.6cm
{\leftline{\tenbf Acknowledgements}}
\vglue0.4cm
We would like to thank T. Banks, D. Kutasov, R. Leigh, M.R.  Plesser, P.
Pouliot, S.  Shenker, M. Strassler and E. Witten for useful discussions.
This work was supported in part by DOE grant \#DE-FG05-90ER40559 and in
part by NSF grant \#PHY92-45317.

\listrefs

\bye